# Band Splitting and Long-lived Carrier Recombination in Ferromagnetic CrSiTe$_3$ Nanosheets

Giriraj Jnawali[*,1], Seyyedesadaf Pournia[1], Eli Zoghlin[2,3], Iraj Abbasian Shojaei[1], Stephen D. Wilson[2,3], Jacob Gayles[4], and Leigh M. Smith[*,1]

[1] Department of Physics, University of Cincinnati, Cincinnati, OH 45221, USA
[2] Materials Department, University of California, Santa Barbara, CA 93106, USA
[3] California NanoSystems Institute, University of California, Santa Barbara, CA 93106, USA
[4] Department of Physics, University of South Florida, Tampa, FL 33620, USA

## Abstract

Magnetic layered ternary chalcogenides hold great promise for future spin-optoelectronic devices in the two-dimensional limit. Understanding how the properties of the materials are impacted by magnetic ordering and the spin-orbit interactions is critically needed information for the development of applications. Ultrafast transient reflectance (TR) and photocurrent (PC) spectroscopies are combined with ab initio density functional theory (DFT) calculations to investigate the band structure and photoresponse of a layered magnetic semiconductor CrSiTe$_3$ (CST) nanosheet in the paramagnetic (PM, 300K) and ferromagnetic (FM, 10 K) phases. We observe both a decrease of the direct bandgap and emergence of a 120 meV splitting of the optical transition when the FM phase is present. DFT band structure calculations suggest that the band modifications are driven by a FM ordering-induced band splitting between the Te $p$ and the Cr $d$ states at the valence and conduction band edges. We find that the majority of carriers photoexcited at the direct gap recombine within picoseconds through defect-mediated recombination, but that 2-3 % of the electrons scatter into indirect conduction band valleys resulting in very long-lived electrons and holes. Those long-lived carriers contribute to the broadband PC response of CST devices that also features indirect absorption. These results provide critical insights into the dynamics and energy landscape of photoexcited electrons and holes, and how they are impacted by spin-ordering effects in layered ferromagnets.

**Significance**

A key to understanding 2D ferromagnetic semiconductors is measuring electrons and holes in their bands and how they are impacted by ferromagnetic ordering. We use mid-IR spectroscopy to show how carriers in direct and indirect valleys are affected by magnetic order. Band structure calculations show energy shifts and splitting of bands are dominated by spin-orbit coupling in the ferromagnetic phase. We show how photoexcited carriers relax by scattering with defects and phonons and reveal a number of channels with dynamics which range over three orders of magnitude. These measurements provide a foundation where both the magnetic ordering and electronic structure can be measured with energetic- and temporal-precision which is critical for development of new physics and spintronic devices.

**Introduction**

The demonstration of magnetism in atomically thin two-dimensional (2D) materials (1-5) has opened up the prospects of new physics and technologies which enable the ability to control the magnetic interactions through the design of heterostructures or the application of electric, magnetic or optical fields (6-12). These depend critically on a thorough understanding of how the spin states, charged carriers in the band structure and their interactions with defects and phonons are affected by the various phases such as ferromagnetic (FM), antiferromagnetic (AF) and paramagnetic (PM) phases which can occur in these materials. For example, calculations of the 2D magnetic semiconductor $CrI_3$ have predicted that the energy gap can change from direct to indirect when tuning the magnetic moment from out-of-plane to in-plane (13). Furthermore, the FM alignment in $CrI_3$ can be optically tuned through the injection of carriers using polarized light (14). Recent ultrafast experiments in $CrSiTe_3$ have shown that selective optical excitation of carriers can increase the FM superexchange interaction between Cr atoms through the Te by a factor of five which potentially can stabilize the FM alignment of the Cr (15). Several Raman experiments have now shown that the symmetry, selection rules, and electron-phonon interactions are changed fundamentally by the magnetic phases within $CrI_3$ and other FM semiconductors (16-18).

Most experiments in 2D magnetic materials have probed the magnetic properties using the magneto-optical Kerr effect (MOKE), SQUID magnetometry, or transport in devices. There have only been a few demonstrations of polarized photoluminescence from electron and hole recombination in $CrI_3$ and $CrBr_3$ thus far, with little information about excited states (19, 20). In this paper we investigate how transient reflectivity can provide new information on both the band structure and carrier dynamics of electrons and holes in quasi-bulk thin nanoflakes of $CrSiTe_3$, a known 2D van der Waals (vdWs) FM semiconductor which has a 33 K Curie temperature ($T_c$) in bulk and 80 K in atomically thin layers (21-24). We show through experiments for both the PM phase at 300 K and the FM phase at 10 K that we can make sensitive measurements of how the band structure is changed for these different magnetic phases, and how measurements of recombination dynamics and spectroscopy provide important hints to the nature of the electronic states and their interactions. Furthermore, these measurements provide an essential basis that can be extended to simultaneous dynamic measurements of both the magnetic phases, and the electrons and holes in the bands. Such measurements are critical to developing a complete understanding of how the states and their interactions can be modified through changing temperature, or the applications of electric or magnetic fields, or the creation of heterostructures.

Magnetic vdWs ternary chalcogenides such as CrSiTe$_3$ (CST) and CrGeTe$_3$ (CGT) are of intense interest as they display magnetic as well as semiconducting character down to thin atomic layers (1, 2, 4, 10-12, 25-28). Recent work has shown CST and CGT devices are very photosensitive (29, 30). Here we use Mid-IR transient reflectance (TR) and photocurrent (PC) spectroscopy to investigate the band structure and dynamics of photoexcited carriers in CST nanosheets in both the PM (300 K) and FM (10 K) phases. Optical transitions are linked to the fundamental direct band gap and mid-gap defects by analyzing the TR spectra and their dynamics. In the FM phase at 10 K, insights into a magnetically driven reduction of the band gap along with a splitting of the valence band (VB) and conduction band (CB) are determined. Comparison with DFT band structure calculations strongly suggest these observations are the result of spin-orbit coupling in the FM phase. Recombination kinetics of photoexcited carriers are measured for both PM and FM phases and the multiexponential decays (both short- and long-lived) provide insights into multiple scattering and recombination processes. We show that $\sim 2-3$ % of the photoexcited electrons scatter into remote indirect CB valleys which yield long-lived carriers which enable photosensitive CST devices over a broad mid-IR range. PC spectroscopy in these devices allows us to disentangle the multiple indirect and direct energy gaps. Finally, the broad implications of these results on potential future applications are discussed.

**RESULTS AND DISCUSSION**

The CST single crystals have a vdW layered structure with each unit cell consisting of three CST layers stacked in an ABC sequence, i.e., alternately arranged Te-Cr-Te and Te-Si-Te sandwiches (Fig. 1a) along the c-axis (31). Each layer exhibits distorted octahedra filled with Si-Si pairs and Cr atoms sandwiched between the top and the bottom Te layers. The top surface in the *ab*-plane exhibits a honeycomb sublattice with Si pairs at the center of the hexagons. The magnetic properties result from the magnetic moments of the Cr atoms, which are aligned parallel to the c-axis (22). Quasi-bulk CST nanosheets (5-10 μm wide and 50-70 nm thick) were mechanically exfoliated from single crystals onto a 300 nm SiO2/Si silicon substrate (see Methods). Samples used for TR measurements were mounted to the cold finger of an optical cryostat and kept in vacuum with a base pressure of $\sim 10^{-5}$ Torr. Other samples were quickly mounted onto a rotational stage for micro-Raman measurements. Raman spectra were acquired at 300 K from different flakes using a 633 nm laser. Raman spectra from these nanosheets show nearly identical spectra (Fig. 1c). Four sharp Raman peaks were observed at 87.2 cm$^{-1}$, 117.5 cm$^{-1}$, and 146.5 cm$^{-1}$, and 214.2 cm$^{-1}$, which coincides with $E_g^1$ (symmetric intra-chain breathing in the basal plane), $E_g^3$, $A_g^3$, and $E_g^4$ (asymmetric stretching along the chain or the c-axis) modes, respectively (32). The Raman spectra change significantly after exposure to air from sharp to broad peaks ( gray curve in Fig. 1c) due to the formation of TeO$_2$ on the surface (32, 33). Typical CST nanosheets are measured using an atomic force microscope (AFM) which show nominal thickness varies from 50-80 nm and lateral size varies 5-10 μm, as shown in Fig. 1b.

**Transient reflectance spectroscopy:** Transient reflectance (TR) measurements are performed on CST nanosheets kept in vacuum using a standard pump-probe setup, as shown schematically in Fig. 1d and also described in Methods. A 1.5 eV pump pulse excites electrons and holes into states well above the direct band gap which changes the reflectance response $\Delta R$ measured by a tunable (0.3 to 1.2 eV) mid-IR probe pulse from a single CST nanosheet. This response is collected as a function of both the probe energy ($E$) and probe delay ($t$) to monitor the scattering of electrons and holes into the different band energy states and defects before they recombine. The $\Delta R$ data are normalized by the reflectance $R_0$ (without pump excitation) at all probe energies $E$, and recorded as the TR signal, $\Delta R/R_0$. Figure 1e,f displays a 2D false-color map of TR response measured at 300 K (PM) and 10 K (FM) where the colors correspond to $\Delta R/R_0$ values for given $E$ (x-axis) and pump-probe delay times $t$ (y-axis). The energy range $0.72 - 0.8$ eV is not accessible by the probe laser. The maps in Fig. 1e,f, exhibit two long-lasting and strong derivative-like TR responses are seen in the low-energy ($\sim 0.4$ eV) and high-energy ($\sim 1.2$ eV) regions. The strong derivative-like responses suggest direct band-edge optical transitions, since indirect optical transitions are too weak to resolve. The response at $\sim 1.2$ eV is assigned to direct band-edge absorption from the top of the VB and the bottom of the CB. We expect that the low-energy response is likely due to the direct absorption from the VB to mid-gap defects. For the nanosheet in the FM ordered magnetic phase at 10 K, we observe a splitting of the ~1.2 eV spectrum into two derivative-like features separated by more than 100 meV. The maximum TR signal has magnitudes on the order of $10^{-3}$ in the low-energy region; the response in the high-energy region is about 20 % lower than at low energies with identical excitation fluencies. This indicates the rapid depopulation of excited electrons at the conduction band-edge due to ultrafast trapping of hot electrons to the mid-gap defects.

**Band-edge electronic structure:** The TR response following pump excitation in semiconductors is caused by the change of the dielectric function $\Delta n$ of the sample associated with a population of non-equilibrium carriers at bands associated with direct optical interband transitions. The interband contributions to $\Delta n$ produce features in the TR spectra around the band-edge which decay as carriers relax to their ground states. Figure 2a shows TR spectra at several probe-pulse delays at 300 K (faint red dots) and 10 K (faint blue dots), respectively. The derivative-like TR features near 0.4 eV and 1.2 eV persist over long time scales at both 10 K and 300 K. Such a derivative-like lineshape is typical of band to band optical transitions (absorption) in semiconductors, which is caused by band-filling effects of photoexcited carriers in the bands which modify the complex dielectric response $\Delta n$ of the material. The lineshape displays little distortion and minimal broadening except for a monotonic decrease of its amplitude as the delay progresses, suggesting weak perturbation of the dielectric response. Under the assumption of weak field perturbation in simple parabolic bands, the lineshape for $\Delta n$ can be assumed to be the third derivative functional form (TDFF) of a Lorentzian functional shape as proposed by Aspnes to analyze TR spectra: (34)

$$\frac{\Delta R}{R_0}(E) \simeq \sum_{j=1}^{n} Re[C_j e^{i\varphi_j}(E - E_j + i\Upsilon_j)^{-m}], \qquad (1)$$

where $C_j$, $\varphi_j$, $E_j$, and $\Upsilon_j$ are the amplitude, phase, transition energy, and the energy broadening parameter of the $j^{th}$ feature, respectively. The exponent $m = 2$ is considered for three-dimensional excitonic transitions. This simple expression cannot provide much physical insight into the nature of the dynamic response of the material but does allow us to extract the transition energies with time without complex many body and Kramers-Kronig analysis. Such calculations are non-trivial due to complex band dispersions near the band edge in the CST band structure. The 300 K PM spectra can be fitted by two resonances ($n = 2$) while the 10 K FM spectra can only be fitted by three resonances ($n = 3$) in which two resonances near 1.2 eV are overlapped but split by ~100 meV. Fits of Eq. (1) to the measured spectra at each delay time are shown as dashed black lines in Fig. 2a. The normalized moduli (absorption profiles) of each resonance are plotted in Fig. 2b. The phase $\vartheta$ and broadening $\Upsilon$ remain identical for each fitting, except at early times (< 2 ps) caused by non-equilibrium hot carrier distributions before thermalization. The delay-dependent transition energies for the 300 K (PM) and 10 K (FM) phases are shown in Fig. 2c. Once thermalized, the transition energies show minimal change over a long delay range, and so they can be considered ground-state optical absorption onset energies. This analysis for the 300 K PM phase exhibits a low-energy transition at $E_d = 0.35 \pm 0.005$ eV and a high-energy transition at $E_1 = 1.18 \pm 0.005$ eV. The 10 K FM phase shows the low energy transition is shifted to $E'_d = 0.37 \pm 0.005$ eV, or blue-shifted by ~ 20meV. Simultaneously the high energy transition is seen to split with two transitions observed at $E'_1 = 1.1 \pm 0.005$ eV and $E''_1 = 1.22 \pm 0.005$ eV. This splitting is thus measured to be $E''_1 - E'_1 \simeq 120 \pm 10$ meV, which can also be seen in the spectral map (Fig. 2e) where double zero-crossings appear over a long delay range. Note that the full high energy spectrum could not be measured at the highest energies which causes some uncertainty in estimating the $E''_1$.

Band structure calculations of bulk CST estimate a fundamental indirect gap at 0.4 eV and the lowest energy direct gap at 1.2 eV as well as multiple indirect gaps in between.(27, 35, 36) From these calculations we can clearly identify the high energy transition as the optical transition between the Te $p$ states at the VB maximum to the empty Cr $d$ states at the CB minimum, which was also observed recently in absorption measurements (37). Since the $E_d$ transition at 0.35 eV is well below the fundamental indirect band gap (0.4 eV) and exhibits as strong as resonance as the high energy transition, we assign it as the direct transition from the VB maximum to mid-gap defect states $E_{def}$. Recent studies (38) have shown that it is energetically favorable to form Cr/Si anti-site defects in CST that form a robust band deep within the energy gap and so it is expected to have a strong TR response. It also was shown recently in CrGeT$_3$ that similar Cr/Ge defects produce a finite density of in-gap states (39). Since absorption into the indirect CB minimum must involve a phonon to conserve momentum, we expect such transitions to be orders of magnitude weaker.

At 10 K, when the nanosheet is in the FM phase, the defect-related transition is slightly blue-shifted by ~ + 20 meV, while the direct gap transition red-shifts to lower energies by ~ − 80 meV and appears at 1.1 eV. A second resonance appears which is 120 meV higher at 1.2 eV. Most semiconductors exhibit a blue shift of the optical gap at low temperatures due to phonon renormalization. We propose that the most likely origin of our observations of the red-shift and splitting of the low-temperature optical transitions is driven by FM ordering induced band splitting below the Curie temperature, $T_c$, as shown schematically in Fig. 2d. For the PM phase at RT, both CB and VB extrema, which are mainly associated with crystal field split Cr $3d$ and Te $5p$ states, respectively, are coupled only by the Coulomb interaction, resulting in a single direct transition within that energy range. With FM ordering, the finite exchange coupling constant and magnetization below $T_c$, each band splits into two sub-bands, a band for spin parallel and one for spin antiparallel to the magnetization, resulting in non-degenerate states in the band structure. This phenomenon can be understood as a first-order energy change due to exchange coupling between charge carriers and spins, as described previously using perturbation theory within an effective mass approximation (40). A deep-level defect should exhibit minimal splitting because of strong localization. While there are four possible transitions between the split states at the VB maximum and CB minimum, we observe only two transitions in the high energy region, perhaps limited by the relatively poor energy resolution available in TR pump-probe measurements. We suggest that the lower energy $E_1'$ transition is the transition between the uppermost VB and lowermost CB, while the higher energy $E_1''$ transition is from the lower spin-split VB to the upper spin-split CB. The defect-related transition at low energy should see only the VB splitting. We observe only one transition $E_d'$ at 10 K which is 20 meV blue-shifted as compared to $E_d$ at 300 K. We assume that there might be a second transition at lower energy but is not visible because it is at lower energy than can be accessed with our laser system. We estimate that the splitting of the VB must be $\Delta_{VB} > 40$ meV because otherwise we would expect to see a second spectral line at low energies. With this assumption we conclude that the CB splitting must be $\Delta_{CB} \sim E_1'' - (E_1' + \Delta_{VB}) < 80$ meV in order to account for the total observed splitting of $\sim 120$ meV.

To understand the physical origin of the observed spectroscopic splittings, we use ab initio density functional theory (DFT) calculations (see Methods) with the GGA exchange-correlation potential to simulate the band structure of the CST with the experimental lattice parameters. The magnetic moment of the unit cell is 2.92 $\mu_B$ with spin-orbit coupling (SOC) and 2.90 $\mu_B$ without SOC. The Cr atoms show an average moment of 3.1 $\mu_B$ with and without SOC primarily due to the Cr $d$ states. As expected, the Si ion shows little to no magnetic moment, while the Te ion shows a moment that is aligned antiparallel to Cr and averages $-0.10$ $\mu_B$ ($-0.11$ $\mu_B$) with (without) SOC due to the Te $p$ states. The moment induced on the Te ion is due to the hybridization of the Cr $d$ and Te $p$ states in the CB. In Fig. 3a, we show the spin-polarized band structure of CST, calculated without SOC, which is consistent with previous studies.(41) The conduction band minimum (CBM) has both minority spin (↓ red) as well as majority spin (↑ blue) bands located along the high symmetry directions $L - \Gamma - K - M$ in the Brillouin zone (Fig. 3a), showing a series of closely linked indirect gaps, respectively. The fundamental direct gaps both with the majority and minority spins are located at the $\Gamma -$ point, in which the gap with minority spin is nearly twice wider than the gap with majority spin. Figure 3b shows the element-selective band structure with SOC included in which the Cr $d$ states (plotted as purple dots) dominate the CB structure while Te $p$ states (plotted with light green dots) dominate the VB edge. Strong $d - p$ hybridization between Cr $d$ and Te $p$ states generates Te $p$ holes with opposite spins at the uppermost VB near the Fermi energy at the $\Gamma$-point and the hybridized Cr $d$ states (i.e., empty Cr $e_g$ states) at the lowermost CB, respectively. The Si contributions to the electronic structures are minimal for both the VBM or CBM, and so it is not included here. Interestingly, we observe a substantial change in the Cr $e_g$ and Te $p$ states around the band edge region.

Analysis of these results allows us to make some conclusions as to the origin and magnitude of the observed 120 meV splitting of the direct band edge lineshape. The DFT calculations without SOC (Fig. 3a) show only splitting of the VB while the direct CB edge is degenerate. This provides a necessary condition that the VB would have to provide the entire 120 meV splitting without including SOC. Since this is three times larger than the ~ 40 meV splitting of VB, estimated by DFT without SOC, this seems unlikely. On the other hand, DFT calculations with SOC show splitting of both the VB and CB, with the VB splitting (Fig. 3d) approximately twice the CB splitting (Fig. 3c). Therefore, it allows us to conclude that the observed 120 meV splitting in the high-energy optical transitions at FM phase originate from the Te $p$ states in the VB with an approximate splitting of $\Delta_{VB}$~ 80 meV (Fig. 3d) and the hybridized Cr $d$ (i.e., Cr $e_g$) states in the CB with SO splitting of $\Delta_{CB}$ ~ 40 meV (Fig. 3c). Despite the underestimation of the energy gap, which is typically expected in DFT, the total splitting of the VB and CB calculated by DFT with SOC (double arrows, Fig. 3c,d) is remarkably consistent with the observed spectroscopic splitting.

**Photocurrent (PC) spectroscopy:** The TR measurements show direct optical transitions at 1.1 eV at the energy gap at zone center and a transition at 0.35 eV which is the transition from the VB to a mid-gap defect state. TR measurements are not sensitive to the indirect CB minima. PC spectroscopy, however, measures the occupation of electrons and holes in all bands directly. It is sensitive to optical absorption of electrons from the zone center VB to the indirect CB valleys through phonon-assisted transitions. PC spectroscopy measurements were made on a single CST nanoflake device, as shown schematically in Fig. 4a (see Methods). The sample is found to be highly conductive (resistivity, $\rho$ ~ 7.5 k$\Omega \cdot$ cm) in the dark at 300 K, as shown by the linear *IV* curve in Fig. 4b. Since the measurements of the mobility in CST (36, 42) vary widely and ranges from 0.02 to 20 cm$^2$/V.s, we can only estimate that the carrier density ranges from $10^{13}$ to $10^{16}$ cm$^{-3}$, which are reasonable carrier densities for an unintentionally doped semiconductor. For PC measurements the mid-IR laser is focused onto the device using a 50 × reflective objective to a ~ 2 − 3 μm spot. With 1030 nm illumination, the slope of the linear *IV* curve increases substantially. The PC response increases linearly with the power of the laser pulse as well as the applied bias (see Fig. 4c) confirming photoinduced carrier generation in CST. From the power dependence we estimate the device sensitivity to be 7.5 μA/W at 1.2 eV (1030 nm). For PC spectroscopy, a transimpedance amplifier is used to measure the current at constant bias as the laser energy is tuned. The laser is mechanically chopped, and the signal is measured using a lock-in amplifier. The magnitude of the PC strongly varies with the photon energy of the incident laser as shown in Fig. 4d. PC spectra are obtained at RT with a constant bias of 0.5 V as the mid-infrared laser is tuned from 0.3 eV to 1.25 eV. These photon energies lie mainly below the direct band gap. Figure 4e shows the PC spectrum plotted on a logarithmic scale, which is normalized by the number of incident photons. An important feature is the rapid increase of the PC signal at 0.37 eV and 0.8 eV. The increase of the PC signal above 0.4 eV is attributed to the excitation of electrons from the zone center VB to the indirect CB valley, which determines the fundamental indirect energy gap. The increase at 0.8 eV is associated with the excitation of electrons from the zone center VB to a higher-lying indirect CB valleys, which are seen in band structure calculations. That PC absorption features are not seen in TR provide strong evidence that these are k-space indirect phonon assisted optical transitions.

**Carrier decay dynamics:** Photoexcited carrier dynamics of these transitions are measured using long-delay time traces of TR at 300 K (PM) and 10 K (FM), as shown in Fig. 5a,b, and the insets. Each time trace displays a multi-exponential decay that occurs at substantially different time scales (ranging over orders of magnitude), which indicates multiple decay channels all temperatures. To quantify the decay process, we analyze each time trace with a tri-exponential decay function, convoluted by the 120 fs wide Gaussian instrument response function, $I_{IRF}$:

$$\frac{\Delta R}{|\Delta R_{\text{peak}}|}(E,t) = I_{IRF} \otimes \left( \sum_{i=1}^{3} A_i \cdot e^{-\frac{t-t_0}{\tau_i}} \right) \qquad (2)$$

where $\Delta R/|\Delta R_{\text{peak}}|$ is the TR signal normalized by the peak magnitude, $t_0$ is time zero, and $A_i$ and $\tau_i$ are the amplitudes and decay time constants of each decay channel, respectively. The decay lifetimes extracted from fits to the data are shown in Fig. 5c. Three distinct dynamic regions are seen in the decay process. The initial fast decay occurs within 2 ps, where the majority of the signal ($\sim 80\%$ of the peak) relaxes. Subsequently, nearly $10-15\%$ of the peak signal decays more slowly within 40 ps, followed by very slow kinetics of residual signal ($2-3\%$ of the peak) that persists over several nanoseconds. The log-log plots of each time trace, as shown in the insets of Fig. 5a,b, further clarify these three distinctive dynamic regions, which also vary with temperature.

The dynamic processes of photoexcitation and relaxation are shown schematically in Fig. 5d. Upon pump excitation (red vertical arrow), hot carriers (hot electrons and holes) $\Delta N$ are generated at high energies in the valence and conduction bands of the CST that causes an abrupt change of the dielectric response $\Delta n$ of the sample which is reflected in the rapid rise of the TR response. Due to the strong electron-phonone coupling strength, $\gamma_{e-ph}$, and the excess energy of the electrons and holes, $\delta E \sim 0.3$ eV (much larger than the optic phonon energy), the photoexcited hot carriers (electrons and holes) should thermalize by phonon emission in the first few hundred fs after photoexcitation, similar to what happens in other systems.(43) However, the majority (~80%) of the TR signal decays extremely rapidly within ~ 1 ps at nearly all probe energies, and so we attribute this rapid decay to Shockley-reed-hall (SRH) interband (nonradiative) recombination mediated by the mid-gap defects (see left vertical arrows in Fig. 5d). This is consistent with the presence of strong mid-gap defects which we have described previously. Thus, the hot electrons rapidly are captured from the CB to these mid-gap defects, which then recombine with holes at the VB edge which causes the abrupt ultrafast decay of the TR response. This first decay constant $\tau_1$ varies between $\sim 1.2 - 2$ ps near 0.35 eV and $\sim 0.8 - 1.5$ ps near 1.1 eV at 10 K (FM) and 300 K (PM), respectively. The slightly slower decay rate at low-energies is due to the screening of electron-phonon interaction by the presence of defects, which is seen in doped semiconductors (44, 45). We note that $\tau_1$ increases by a factor of two at 300 K at both transition regions (see Table 1). This is not seen in typical semiconductors since the exchange rate with the lattice usually increases rapidly at higher lattice temperatures. Unlike in metals, the electron-phonon coupling constant in semiconductors is usually temperature independent. The faster decay rate at 10 K suggests additional scattering channels active in the FM phase, which may be reflected in the appearance of the different transitions, i.e., $E_d'$, $E_1'$, and $E_1''$ at 10 K.

After this initial ultrafast decay, a slightly slower decay constant $\tau_2$ appears which we attribute to slow defects (see right vertical arrows in Fig. 5d). Similar slower recombination processes, which follow an initial rapid process, has been seen to occur in materials that have different types of defects with different capture rates.(46-48) The CST samples are known to have several anti-site defects at the mid-gap. The measured $\tau_2$ near the $E_1$ region is $\sim 40$ ps but is reduced to $\sim 20$ ps near the $E_d$ region, which is expected due to high probability of direct recombination to VB holes. The slower decay $\tau_2$ is the same at 10 K and 300 K (Fig. 5c). This may reflect that the slower defect does not couple as strongly with phonons, and so decay may be mediated through an Auger process (49, 50).

The third and final decay time of the remaining signal ($\sim 1 - 2$ %) is very long. We estimate this time to be $\sim 5$ ns at low and high energies (see Fig. 5c), but it is likely even longer due to limitations of our probe delay system. This long-lived residual signal might be due to slow release of defect-trapped carriers but is more likely due to scattering of electrons into remote indirect CB valleys which cannot decay with VB holes without the mediation of a phonon to conserve momentum. This is supported by the fact that this weak residual signal is observed at all energies. TR measurements are not sensitive to electrons trapped in the indirect CB valleys, but is sensitive to the holes remaining at zone center. This scattering process is shown by dashed blue curved lines in Fig. 5d.

Comparison of the PC measurements with decay dynamics reveals some basic physical insights about electronic correlation in the CST sample. It has been widely discussed that strong Coulomb interactions from the narrow Cr $d$ bands dominates over short-range magnetic correlation above $T_c$ that favors Mott-type insulating phase in CST, which is further dominated by the FM superexchange interaction below $T_c$ (41). Although the nature of the Mott gap depends on electronic interactions and short-range magnetic correlations, which have not been determined experimentally, we expect that the strong electronic correlations with high charge density at the Fermi energy, would result in a fast decay of photoexcited carriers (51, 52). This would substantially suppress PC at room temperature. Since we observe long-lived photocarriers and strong PC response, these results provide strong evidence that CST behaves like an ordinary PM semiconductor above $T_c$ at 300 K and an FM semiconductor below $T_c$ at low temperature. These results show the great potential of CST-based devices for both optoelectronic as well as spintronic devices.

Now we return to the dynamic spectroscopic shifts in TR spectra which are observed at early times (< 2 ps). Such shifts in nonmagnetic semiconductors are usually related to band-gap-renormalization (BGR) due to many-body interactions of the electrons and holes. However, in a magnetic semiconductor, a dynamic change in the magnetization may potentially provide an additional shift in the band structure. The dynamic shift for $E_d$ (~ 0.35 eV) at 300 K is ~$18 \pm 2$ meV and for $E_d'$ at 10 K it is ~$25 \pm 2$ meV. In contrast, there is almost no dynamic energy shift (~$3 \pm 2$ meV) observed for $E_1$ at 300 K. At 10 K, however, an additional dynamic red-shift is observed for $E_1'$, ($\Delta_g$~$23 \pm 2$ meV, see Fig. 2c), which is added to the late time red-shift of $\sim -80$ meV. The upper state $E_1''$ exhibits a small blue-shift at early times. By comparison with the time decays, we conclude that the dynamic energy shifts are associated with the electron occupation of the CB at zone center. These electrons decay rapidly through defect mediated SRH recombination, and the observed energy shifts scale with this carrier decay. Since $E_1$ at 300 K shifts quite smaller than for $E_1'$ at 10 K, it is possible that this dynamic energy shift may result from interactions with the magnetic phase of the material. Virtual hopping from half-filled Cr $t_{2g}$ orbitals to the empty Cr $e_g$ orbitals gives rise to FM exchange coupling in CST (53-55), and so additional electrons at the Cr $e_g$ orbitals through pump excitation might modify the FM exchange interaction. Indeed, recent ultrafast reflectivity measurements have shown strong enhancement of superexchange coupling between Cr atoms through Te with optical excitation around 1 eV (15). Therefore, the TR results discussed here not only demonstrates direct probe of band structure but also paves the way for the unique possibility of transiently manipulating the electronic structure by altering the exchange interaction via selective optical excitations in magnetic materials (56, 57). Similar measurements would provide comparisons in various other vdWs magnets to CST.

In conclusion, we have used transient reflectance (TR) to measure the photoexcited carrier dynamics and band structure of CST for both the PM phase at 300 K and the FM phase at 10 K. At 300 K, a strong direct optical transition is observed at the direct band gap at ~ 1.2 eV and also a direct optical transition near 0.37 eV between the top of the VB and a deep level defect state. In contrast, PC spectra in a single CST nanoflake device show features at the fundamental indirect gap at ~ 0.4 eV as well as a second higher-lying indirect CB at ~ 0.8 eV. The TR time-traces show that the lifetime of electrons at the direct CB decay rapidly within a few picoseconds (~2 ps) due to SRH recombination through the defect state. However, longer-lived trapped electrons are seen in a secondary defect state which decay within ~ 40 ps and extremely long-lived electrons are seen in the indirect CB edge, which persist for over 5 ns. TR spectroscopy shows that in the FM state, the VB and CB extrema at the zone center are split by ~ 80 meV and ~ 40 meV, respectively, resulting in two optical transitions observed near 1.2 eV separated by ~ 120 meV. Detailed DFT calculations, which include SOC, show that these splittings result from the strong contribution of $d-p$ hybridization of Cr $d$ and Te $p$ states at the conduction and valence band edges at zone center. These results show that transient optical spectroscopy can provide insight into how the band structure changes in 2D materials as a result of order/disorder transitions of the magnetic states, and the dynamics of photoexcited carriers can provide insights into their interactions and recombination processes. We believe that these results show a way of moving forward towards comprehensive experiments where both the magnetic phases and electronic states in the bands can be simultaneously probed optically to gauge the impact of applied fields or fabrication of heterostructures.

## Methods

**Sample preparation:** CST single crystals were grown using a self-flux technique (21). The constituent materials Cr (99.999%, Alfa Aesar), Si (99.999%, Alfa Aesar), and Te (99.9999%, Alfa Aesar) were loaded into a 5 ml alumina crucible (reaction vessel) with a molar ratio of 1:2:6. The reaction vessel was then placed inside a quartz tube with a secondary catch crucible filled with quartz wool immediately above the catch crucible. Care was taken to minimize exposure of the constituent elements to air, and the reaction preparation was performed in an Ar-filled glovebox. A valve was attached to the ampoule, which was then removed from the glovebox, evacuated ($5 \times 10^{-5}$ mbar) and backfilled with Ar several times before sealing under approximately 1/3rd atmosphere of Ar. The sealed ampoule was heated to 1150°C at ~ 2°C/min, followed by a 16 h dwell at temperature, and then cooled to 700°C at a rate of 3°C/h. The ampoule was removed from the furnace at 700 °C and quickly centrifuged to remove excess Te flux. Plate-like crystals up to 5 mm thick with flat and highly reflective surfaces were then removed from the reaction crucible.

Thin flakes (nanosheets) were prepared from these single crystals by using a standard mechanical exfoliation technique on an oxidized and low-doped silicon substrate (300 nm $SiO_2$/Si). The samples were then immediately mounted onto the cold finger of a liquid helium cooled cryostat (Janis ST-500) for optical measurements, which operates between 8 K and 300 K by maintaining the sample under high vacuum conditions (pressure $< 1 \times 10^{-4}$ Torr).

**Sample characterization**. Surface morphology and thickness of CST nanoflakes were measured by using atomic force microscope (AFM). Measurements were conducted in air on a Veeco AFM using gold (Au) coated platinum (Pt) cantilevers operating in tapping mode. Raman spectroscopy was applied to characterize film quality and possible degradation of sample by oxidation. The measurements were performed using a Dilor triple grating spectrometer coupled to a microscope with 632.8 nm laser excitation in the backscattering configuration under ambient conditions. A 100 × objective was used to focus the incident beam and collect the scattered signal and then dispersed by 1800 g/mm grating after the resonant scattered laser light is removed using a double subtractive mode spectrometer. The Raman scattered signal is measured with a spectral resolution of 1 cm$^{-1}$. The lowest frequency observable is 50 cm$^{-1}$. Laser power was maintained around 100 µW during the measurements to minimize laser-heating effect in our samples.

**Transient reflectance spectroscopy:** Ultrafast transient reflectance spectroscopy measurements were performed using a Ti:sapphire oscillator which produces 150 fs pump pulses at a central wavelength of 800 nm with 80 MHz repetition rate and an average power of 4 W. The majority (80 %) of the laser beam was used to pump an optical parametric oscillator (OPO), which generates variable wavelength signal beam (0.8 – 1.2 eV) and an idler beam (0.3 – 0.72 eV). All measurements were performed using photoexcitation at 820 nm with a photon energy of $\hbar\omega_p = 1.51$ eV and probing with both signal and idler outputs, respectively. The probe pulses are delayed in time with respect to the pump pulses by using a motorized linear translation stage. The pump and probe beams are spatially overlapped within a focused spot size ~ 2 µm in diameter on a single CST flake using a 40 × objective with the help of CCD camera and the reflected beams are directed to the liquid nitrogen cooled InSb detector. The pump beam is filtered out during the TR measurements using a long pass filter. The signal is collected with a lock-in amplifier phase-locked to an optical chopper that modulates the pump beam at a frequency of 1 kHz. An average pump fluence of ~ 0.3 mJ/cm$^2$, which is still well below the damage threshold in bulk CST, was used throughout the experiments, while the probe fluence was kept nearly one order of magnitude lower.

**Device fabrication:** Rectangular shaped CST nanosheet is prepared on Si substrates (with 300 nm SiO$_2$ film) by mechanical exfoliation for the fabrication of two-terminal devices (Fig. 4a). Large metallic contact pads of 300 nm Aluminium (Al) on 20 nm Titanium (Ti) are fabricated on either side of the flakes using a standard photolithographic process. Before metal deposition, the sample was immersed in a solution of HF (1)∶H$_2$O(7) for a few seconds to remove oxides and photoresist contaminants, which allow to achieve an ohmic contact between the flake and the metal contacts. The sample is mounted onto the cold finger of a He$^4$ constant-flow cryostat (Janis ST-500), which allows cooling of the sample down in vacuum to 10 K. Prior to PC spectroscopy measurements, all devices are checked by acquiring *IV* curves in dark.

**Band structure calculations:** First Principle calculations with Vienna ab initio Simulation Package (VASP) (58, 59) were carried out on the conventional unit cell of CST which consisted of 3 primitive unit cells. We make use of the generalized gradient approximation (GGA) exchange correlation potential. The self-consistent calculations are converged on an $8 \times 8 \times 4$ k-point mesh with an energy cutoff of 245 eV. The FM is shown to be the ground state in comparison to a collinear AF arrangement of spins and a random orientation of spins where the total moment is zero.

## Acknowledgments

L.M.S. acknowledges the financial support of the NSF through grants DMR 1507844, DMR 1531373 and ECCS 1509706. J.G. acknowledges the financial support of USF. S.D.W and E.Z. acknowledge support from ARO Award No. W911NF-16-1-0361.

## Author contributions

G.J. and L.M.S. conceived the study. E.Z. and S.D.W. synthesized CrSiTe$_3$ samples. G.J. characterized the samples, performed TRS measurements, data analysis, figure planning, and draft preparation. J.G. performed ab initio DFT calculations. I.A.S. helped in the micro-Raman measurements. S.P. fabricated and characterized CST devices. S.P. and G.J. performed PC experiments. G.J. and L.M.S. wrote the manuscript. All coauthors contributed to the discussion of results and commented on the manuscript.

## Corresponding authors

*E-mail: giriraj.jnawali@uc.edu (G.J.)
*E-mail: leigh.smith@uc.edu (L.M.S.)

## Additional information

The authors declare no competing financial interests.

Table 1. Amplitudes and average decay constants of tri-exponential fit for pump-probe TR traces of CST samples at two different temperatures (T) and probe energies (E).

| T (K) | E (eV) | $A_1$ (%) | $\tau_1$ (ps) | $A_2$ (%) | $\tau_2$ (ps) | $A_3$ (%) | $\tau_3$ (ps) |
|---|---|---|---|---|---|---|---|
| 300 | 1.2 | 83 | 1.5 | 13 | 37.5 | 4 | 5000 |
|  | 0.4 | 70 | 2.0 | 22 | 17.5 | 8 | 5000 |
| 10 | 1.2 | 89 | 0.8 | 6 | 36.5 | 5 | 5000 |
|  | 0.4 | 79 | 1.2 | 12 | 19.1 | 9 | 5000 |

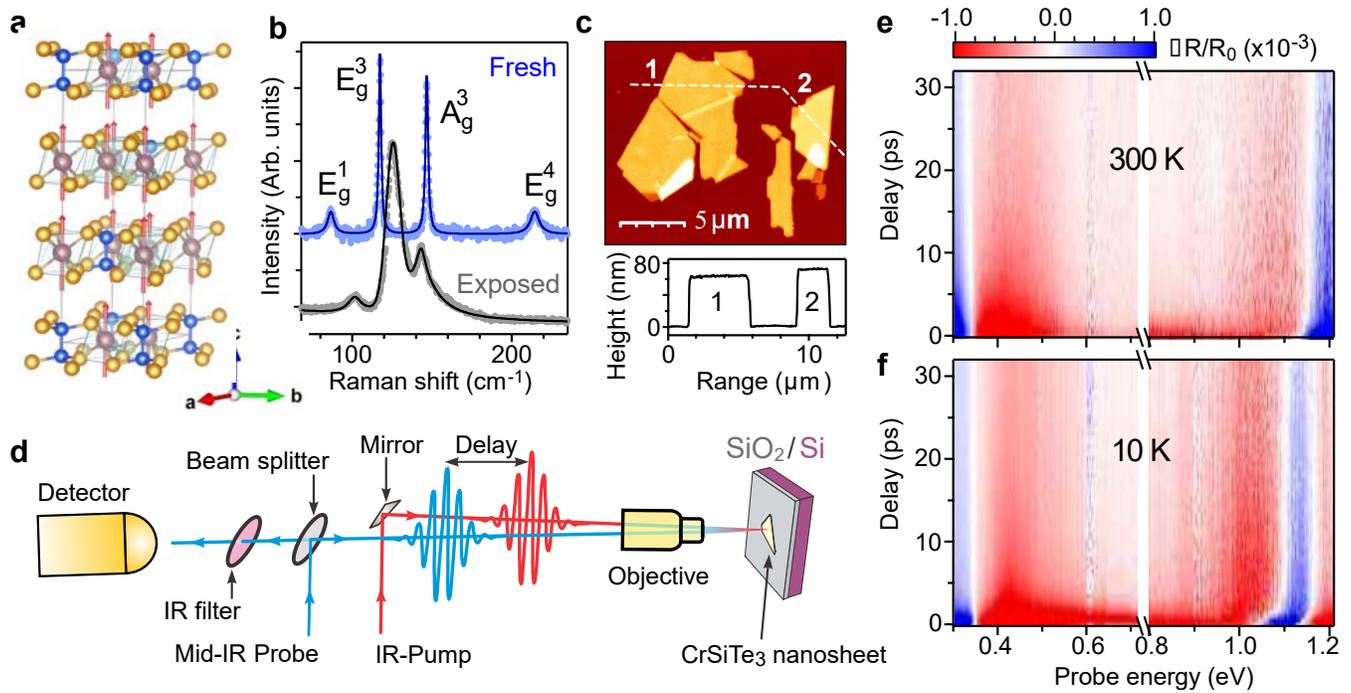

**Fig. 1** Structure and morphology of a CST nanosheet and TR measurement scheme. **a** Crystal structure of CST, showing ABC stacking of 2D atomic layers along the c-axis. Cr Spin moment is also aligned along the c-axis (red arrow). **b** Typical Raman spectra measured on a CST nanosheet immediately after exfoliation (blue line) and after long-exposure to air (gray line). Measurements were performed in air with 633 nm laser. **c** Tapping-mode AFM image (upper panel) and height profiles of CST nanosheets exfoliated onto SiO$_2$/Si substrate. Step height profile (lower panel) shows thickness of each flake (∼ 60 − 70 nm). **d** Pump-probe TR measurement setup. The CST nanosheet is optically excited by a 1.5 eV pump pulse with wavelength tunable and delayed probe pulses. **e, f** Pseudo-color maps (spectral and temporal) of TR response at 300 K and 10 K. The TR signal in the range 0.3-0.72 eV is scaled by 1.5 to optimize the contrast. Derivative-like features associated with band-to-band optical transitions are seen at low- and high-energy spectral regions.

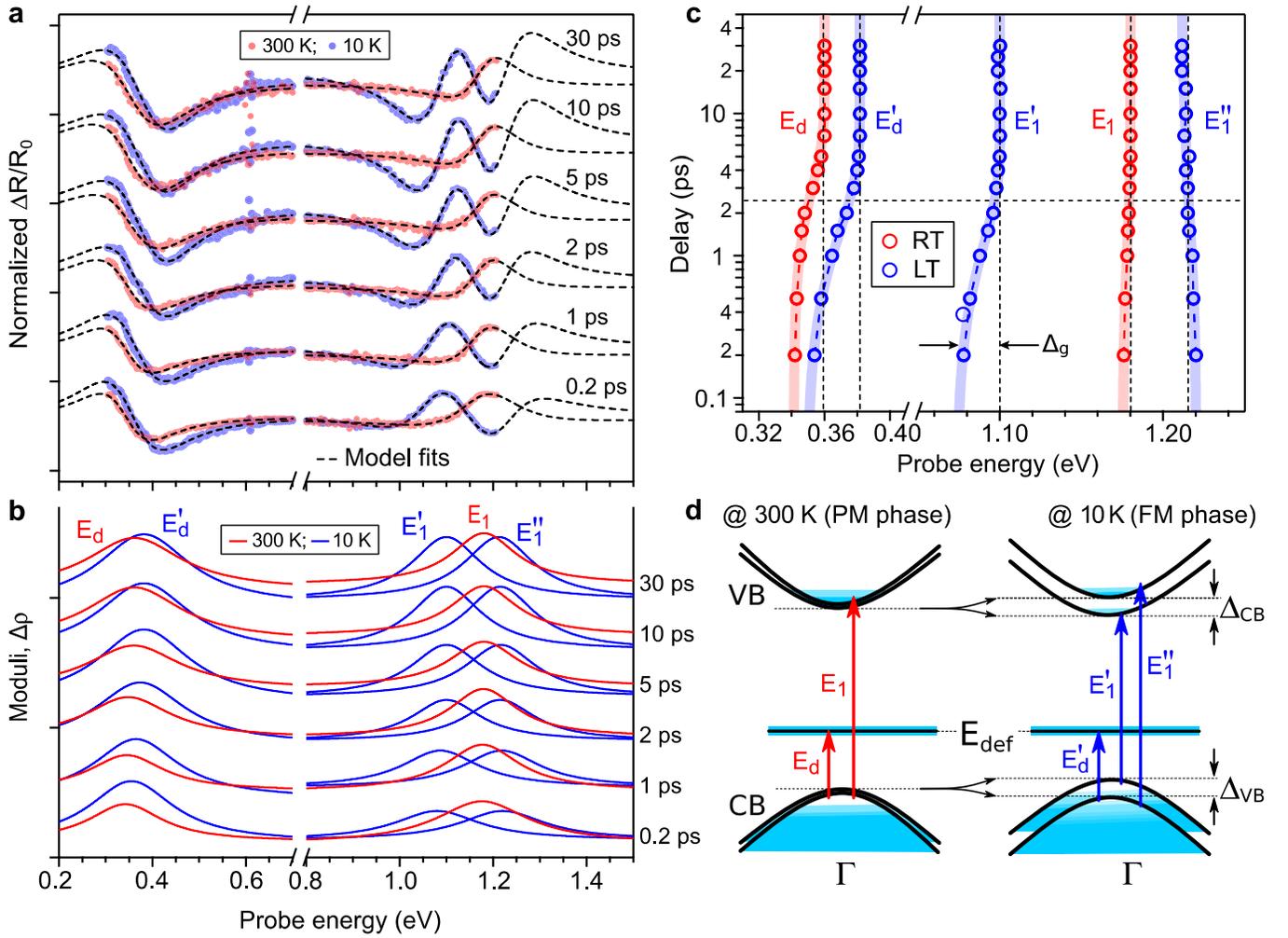

**Fig. 2** Dynamics of band-edge optical transitions in CST nanosheets. **a** TR spectra recorded at different pump-probe delay times at 300 K (light red spheres) and 10 K (light blue spheres). Dashed black lines show fits to the data. **b** Calculated moduli (absorption) of each corresponding fit at 300 K (red) and 10 K (blue), vertically offset for clarity. Both model fits and their moduli show features associated with optical transitions in CST bulk band structure. **c** Transition energies $E_d$, $E_1$, $E'_d$, $E'_1$, and $E_1^{\parallel}$ extracted from the fits and plotted as a function of delay for both temperatures. Ultrafast dynamic shift $\Delta_g$ is indicated. **d** Assignment of possible optical transitions in the band structure of CST (schematic) at 300 K (PM) and 10 K (FM). Schematic illustration of magnetic-order-induced splittings of VB and CB edges at 10 K. Arrows indicate optical transitions at 300 K (red, PM) and 10 K (blue, FM).

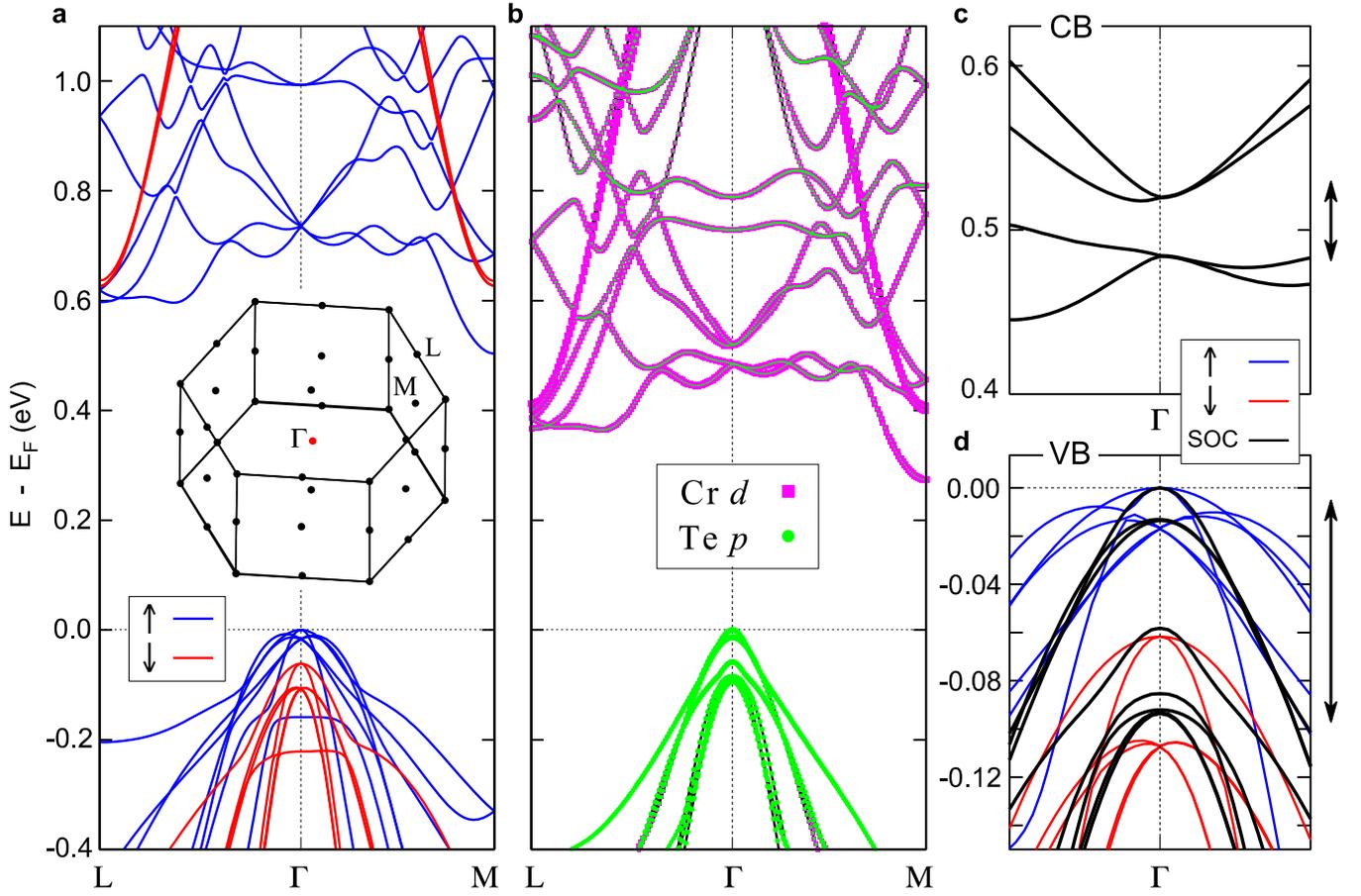

**Fig. 3** Bulk band structure of CST near $\Gamma$- point. **a** Spin-polarized band structure without including SOC. Blue and red curves indicate the majority ($\uparrow$) and minority ($\downarrow$) spin states, respectively. First Brillouin zone of CST with high symmetry points ($\Gamma$, $L$, and $M$) indicated in the inset. **b** Element selective band structure with SOC included (black lines are guide to the eye). The Cr $p$ states (light green dots) are dominant around the VB edge and Cr $d$ states (i.e. Cr $e_g$ states) (light blue dots) are dominant around the CB edge. Both VB and CB show splitting at the $\Gamma$- point. **c, d** Close-up of CB and VB, respectively. Black lines represent the bands with SOC included (combining all element selective bands) while blue and red lines are the spin polarized bands from **a**. Approximate VB and CB splittings with SOC shown by double arrows.

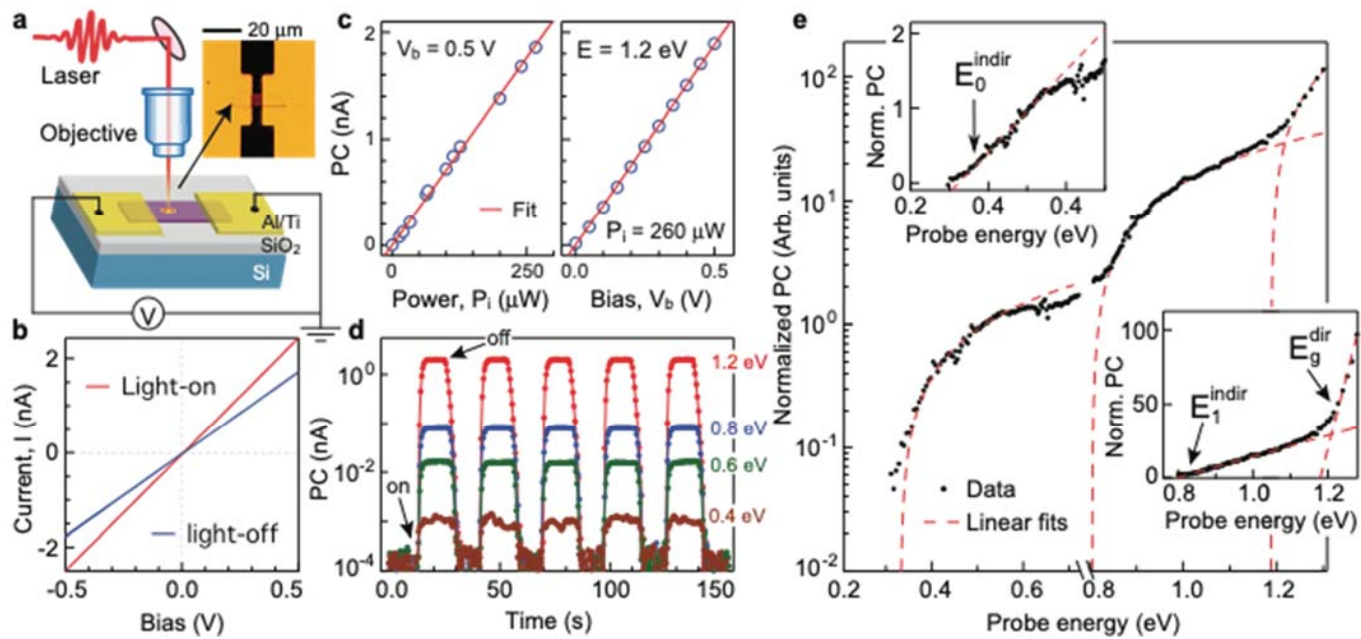

**Fig. 4** PC response of a CST nanosheet. **a** Schematic view of CST device and PC measurement scheme. Optical image (top view) of CST device. **b** *IV*-characteristics of the device in dark and 1030 nm light, showing a clear PC response (increased conductance). **c** PC response versus incident laser power ($P_i$) and applied bias ($V_b$). Red lines are linear fits. **d** Excitation energy $E$ dependent PC response of the device during repeated on/off cycles of light illumination. The PC signal decreases at lower photon energies. **e** PC spectrum (normalized by incident photons) measured by varying wavelength of incident laser. Dashed red lines are linear fits around band-edge regions. Insets show linear expanded regions, displaying onset of absorptions for different transitions.

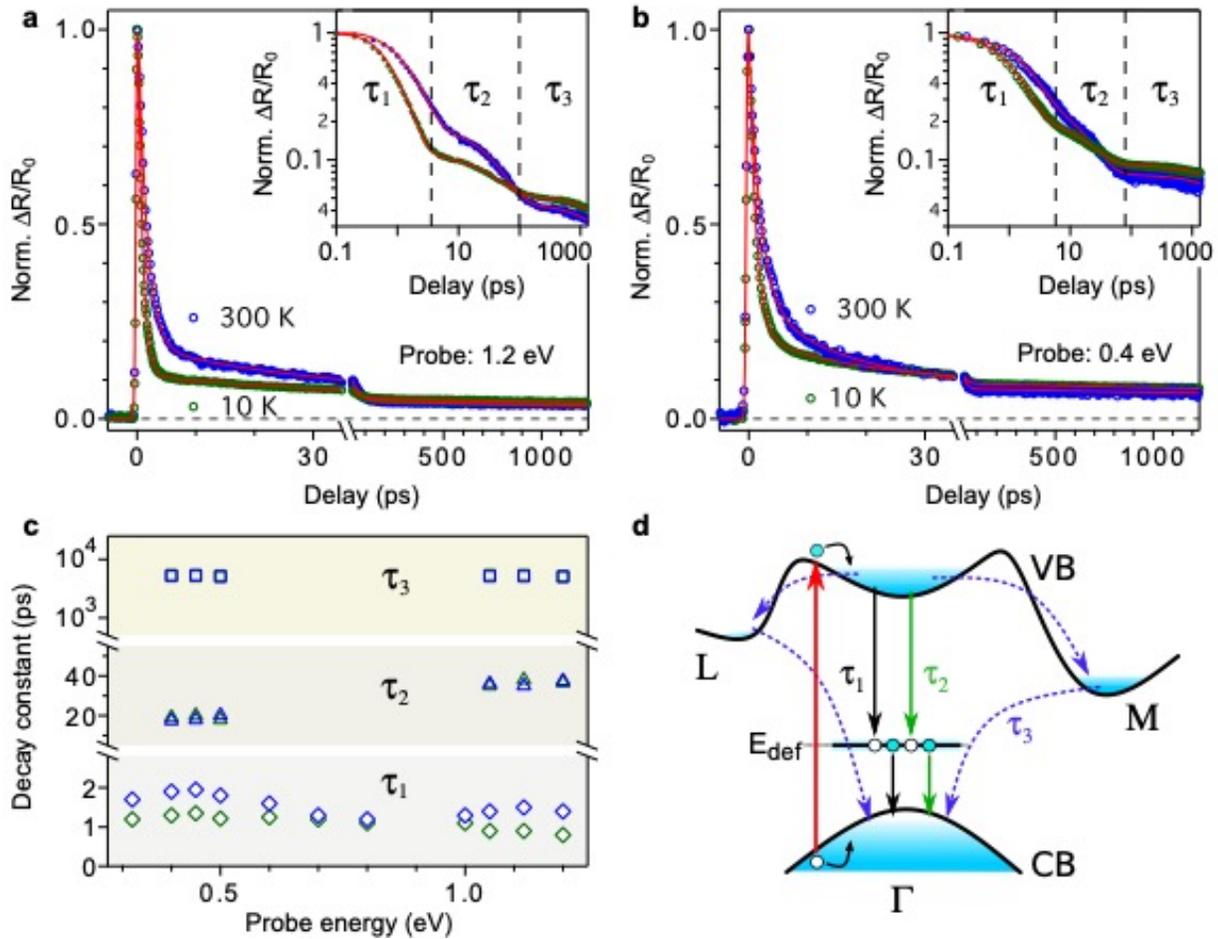

**Fig. 5** Carrier decay dynamics and recombination processes in CST nanosheets. **a, b** TR time decays near high- and low-energy transitions at 300 K and 10 K. Multi-exponential decay dynamics are fit with three dominant decay channels (seen more clearly in insets) using fits to tri-exponential function (Eq. 1) with decay time constants of $\tau_1$, $\tau_2$, and $\tau_3$. **c** Plots of decay constants with energy. **d** Schematic showing of relaxation channels observed in the transient response. Initial fast decay $\tau_1$ corresponds to the intraband cooling of hot carriers by electron-phonon coupling (curved arrows) and subsequent SRH recombination mediated by mid-gap defects at $E_{def}$ in CST band structure (black vertical arrows). The slower kinetics $\tau_2$ is attributed to the recombination by slow defects (green vertical arrows) through an Auger process. Final long-lived kinetics is caused by band-to-band recombination of electrons scattered into indirect valleys that recombine slowly over several nanoseconds (blue dashed arrows) through interaction with phonons to conserve momentum.